\documentclass[10pt,conference]{IEEEtran}
\IEEEoverridecommandlockouts
\usepackage[normalem]{ulem}
\usepackage{algorithm}
\usepackage{braket}
\usepackage[noend]{algpseudocode}
\usepackage{amssymb}
\usepackage{xcolor}

\usepackage[top=1in, bottom=1in, left=1in, right=1in]{geometry}
\usepackage{graphicx,amsmath,amsthm,cite,nicefrac,ifthen,amsfonts}
\usepackage{subcaption}
\usepackage{wrapfig}
\newtheorem{remark}{Remark}

\newcommand{\Znon}{\mathbb{Z}^{\ge0}}

\newcommand{\getCurrentSectionNumber}{%
  \ifnum\c@section=0 %
  \thechapter
  \else
  \ifnum\c@subsection=0 %
  \thesection
  \else
  \ifnum\c@subsubsection=0 %
  \thesubsection
  \else
  \thesubsubsection
  \fi
  \fi
  \fi
}

\theoremstyle{definition}
\newtheorem{thm}{Theorem}[section]
\theoremstyle{definition}

\theoremstyle{definition}

\makeatletter
\def\tagform@#1{\maketag@@@{\bfseries(\ignorespaces#1\unskip\@@italiccorr)}}
\renewcommand{\eqref}[1]{\textup{{\normalfont(\ref{#1}}\normalfont)}}
\makeatother
\begin{document}

\title{Scalable Scheduling Policies for Quantum Satellite Networks}
%\title{Scheduling Policies for Quantum Satellite Networks with Mega-constellations}
\author{
\IEEEauthorblockN{Albert Williams, Nitish K. Panigrahy, Andrew McGregor, and Don Towsley}
\IEEEauthorblockA{University of Massachusetts Amherst, USA. \\
%\IEEEauthorrefmark{2}Yale University, USA. \\
Email: \{abwillia, nitish, mcgregor, towsley\}@cs.umass.edu}}
%\IEEEauthorrefmark{1}

\maketitle

\begin{abstract}
As Low Earth Orbit (LEO) satellite mega constellations continue to be deployed for satellite internet and recent successful experiments in satellite-based quantum entanglement distribution emerge, a natural question arises: \emph{How should we coordinate transmissions and design scalable scheduling policies for a quantum satellite internet?} In this work, we consider the problem of transmission scheduling in quantum satellite networks subject to resource constraints at the satellites and ground stations. We show that the most general problem of assigning satellites to ground station pairs for entanglement distribution is NP-hard. We then propose four heuristic algorithms and evaluate their performance for Starlink mega constellation under various amount of resources and placements of the ground stations. We find that the maximum number of receivers necessary per ground station grows very slowly with the total number of deployed ground stations. Our proposed algorithms, leveraging optimal weighted b-matching and the global greedy heuristic, outperform others in entanglement distribution rate, entanglement fidelity, and handover cost metrics. While we develop these scheduling algorithms, we have also designed a software system to simulate, visualize, and evaluate satellite mega-constellations for entanglement distribution. %Our findings provide valuable insights on the design of future quantum satellite networks.
%We find that the median number of receivers necessary per ground station grows very slowly with the total number of deployed ground stations, and that an assignment algorithm more nuanced than a greedy assignment is needed to approach optimal. 

\end{abstract}

%\section{Abstract}

\section{Introduction}
%{\bf Importance of satellites in QN}\\
Long distant entanglement is a key resource for a variety of distributed quantum applications, such as quantum key distribution (QKD) \cite{PhysRevLett.67.661}, quantum sensing \cite{Degen_2017}, and quantum teleportation \cite{Ma_2012}. However, superposition and entanglements are susceptible to environmental noise and losses. In order to distribute entanglements between distant users, ongoing efforts focus on mitigating noise and increase robustness. Terrestrial entanglement distribution invariably suffers loss that decays exponentially  as a function of distance \cite{Pirandola_2017} and must make use of quantum repeaters \cite{D_r_1999} for long distance distribution. Locally created entangled particles can, instead, travel through free space and incur loss that is only quadratic in the distance travelled \cite{panigrahy2022optimal, Lu2022}. This makes satellite based entanglement distribution \cite{Gundogan2021,Lu2022, DeForgesdeParny2023, Khatri21:Spooky} an attractive alternative to long distance terrestrial links with quantum repeaters.

%{\bf Why LEO?}\\
Satellite based entanglement distribution, however, has its own set of challenges. The rate and fidelity of entanglement distribution through free space depend heavily on  atmospheric conditions such as cloud cover, weather, and time of day. Additionally, one needs to carefully select the optimal altitude for satellite deployment. A higher orbit incurs greater loss, but offers longer contact periods. Conversely, a lower orbit experiences lower loss, but is more dynamic with shorter contact periods. For example, Geostationary (GEO) satellites orbit at  altitudes over $35,000$ kilometers, a range where very few entanglements ($\sim 10^{-4}$) propagate through free space, even though the decay rates are only quadratic. For Low Earth Orbit (LEO) satellites, even in the best case, a single satellite rarely stays within line of sight of a given ground station for more than five minutes. However, the shorter contact periods can be offset by deploying multiple satellites in constellations, ensuring a few of them remain visible to the ground stations most of the time.

Recently, there has been a renewed interest in the deployment of LEO satellite constellations for global scale broad band connectivity \cite{Spacex,Telesat,Amazon}. These constellations, also known as mega constellations,  consist of hundreds to thousands of low cost communication satellites orbiting around the earth at low-orbits. For example, SpaceX has already deployed $6000$ satellites into its Starlink constellation \cite{Spacex}. While these satellites aim to provide conventional classical data communication using RF based satellite-to-ground and Laser based inter satellite links, their potential for enabling photonic entanglement distribution has yet to be explored. Such mega constellations hold promise for significantly improving the rate of satellite based entanglement distribution while providing this service on a global scale.

In a satellite based entanglement distribution, entanglements are distributed to ground station pairs which may be located far from each other. First, an entangled pair of qubits is generated at a source  located on the satellite. Subsequently, these qubits are transmitted to each of the ground stations in the pair through free space. Unlike classical satellite communication, for successful entanglement distribution, the satellite must be simultaneously visible to both the ground stations. We refer to this as the \emph{pair visibility requirement}.

A key aspect of mega constellations is the capability of multiple satellites to maintain visibility with ground stations. Similarly, multiple ground station pairs can be visible to a single satellite at any point of time. Therefore to optimize the operation of such a set-up, i.e. a constellation of satellites serving entanglements to a set of ground station pairs, one needs to solve a scheduling problem - which satellite to distribute entanglements to which ground station pairs and when to do it. This scheduling process must align with the resource constraints present at both satellites and ground stations. We refer to the scheduling problem as the \emph{Quantum Satellite Scheduling Problem} (QSSP). One of the contributions of this work is to devise scalable solutions to QSSP and subsequently evaluate these solutions in the context of satellite mega constellations.

%Also highlight how quantum version of the problem is different and difficult. 
QSSP has been investigated by Panigrahy et. al. \cite{panigrahy2022optimal} with an emphasis on formulating an optimal schedule to maximize the aggregate entanglement distribution rate across all ground station pairs. The authors modeled QSSP as an Integer Linear Program (ILP). While the ILP solution works well for a modest number of ground station pairs and satellites, it does not scale when confronted with a  large number of ground station pairs and mega-constellations. In our work, we prove that the general version of QSSP is NP-hard, by reducing it from a known NP-hard problem. Our focus in this work is on a broader setup where we identify essential features of a scalable and efficient solution to QSSP. In particular, we summarize the major contributions of our work as follows.

%We specifically analyze general-purpose entanglement distribution protocols, which, unlike the quantum keys at trusted intermediate nodes in the protocols studied in \cite{Vergoossen_2020}, cannot be stored as classical information for any length of time. 

% We show that the most general problem of assigning satellites to ground station pairs for entanglement distribution is NP-hard. We then propose four heuristic algorithms and evaluate their performance for entanglement distribution using Starlink LEO satellite mega constellation under various placements and amount of resources of the ground stations. We find that the maximum number of receivers necessary per ground station grows very slowly with the total number of deployed ground stations. Our proposed algorithm based on optimal weighted b-matching outperforms others in entanglement distribution rate, entanglement fidelity, and handover cost metrics. While we develop these scheduling algorithms, we have also designed a software system to simulate, visualize, examine, and evaluate satellite mega-constellations for entanglement distribution. Our findings provide valuable insights on the design of future quantum satellite networks.
\begin{itemize}
    \item We show that the most general version of QSSP is NP-hard by reducing it from a known NP-hard problem, the 3-dimensional matching problem. 
    \item We introduce four heuristic algorithms for QSSP and assess their performance for Starlink mega-constellation across diverse scenarios, including various placements and resource levels at ground stations, while factoring in practical considerations such as time-of-day/year and realistic entanglement generation sources. Our results demonstrate that algorithms based on optimal weighted b-matching and global greedy heuristics outperform others in different performance metrics.
    \item We build a software system for simulation, visualization, and evaluation of satellite mega-constellations for entanglement distribution.
\end{itemize}

%RF to telecom discussion - 
%In this work, we consider LEO satellite networks for entanglement distribution.
%, the quality of the connection and the availability of other connections changing all the while.
%{\bf Why scheduling?}\\
%multiple satellite visibility 
%For broader coverage and to serve numerous ground station pairs, need to deploy a constellation of satellites. 
%In this work we examine the problem of assigning satellite links to ground station pairs. 

\section{Preliminaries}\label{preliminaries}
%\subsection{Modeling assumptions}
%A satellite-based quantum internet can take a variety of forms. Here we describe the specific form we examine in this work and specify the factors we are evaluating.
In this Section, we present the system model used in the remainder of the paper.
\subsection{Satellites and Ground Stations}
\begin{figure}
    \centering
    \includegraphics[width=0.4\textwidth]{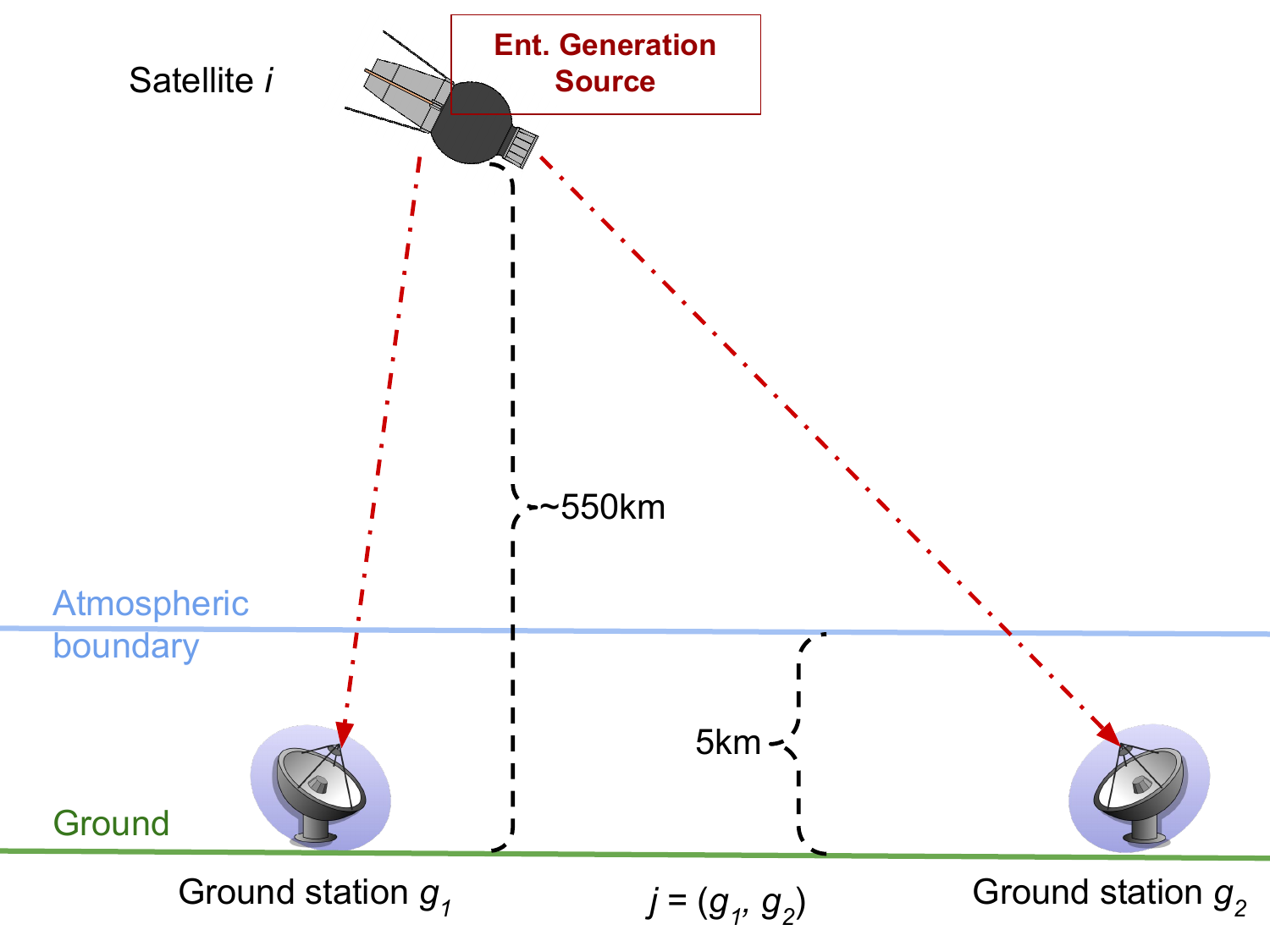}
    \caption{The Dual Downlink architecture for entanglement distribution.}
    \label{fig:dual_downlink}
\end{figure}
\begin{table}[]
\begin{tabular}{l|p{6cm}}
\textbf{Symbol} & \textbf{Description}                                                                     \\ \hline
$S$             & set of satellites                                                                        \\
$T_i$           & ground station pairs satellite $i$ can connect \\
$G$             & set of ground stations                                                                   \\
$R_g$           & receivers on ground station $g \in G$                                                    \\
$F$             & set of ground station pairs $\left(F \subset G \times G\right)$                          \\
$L_j$           & maximum connections between $g_1$ and $g_2$,\newline$j = (g_1, g_2)$                      \\
$w_{ij}(t)$        & value of connecting $g_1$, $g_2$ with $i$, $j = (g_1, g_2)$\\
$x_{ij}(t)$ & decision variable: number of connections $i$ makes between $g_1$ and $g_2$, $j=(g_1, g_2)$, $x_{ij} \in \mathbb{Z}^{\ge0}$
\end{tabular}
\caption{Summary of notation}
\label{table:notation}
\end{table}

Let $S$ denote the set of satellites in the constellation. For each satellite $i \in S$, let $T_i$ denote the number of ground station pairs that satellite $i$ can connect under ideal circumstances. Because each ground station pair comprises two ground stations, $T_i$ is less than the number of transmitters on satellite $i$. The exact relationship between $T_i$ and the number of transmitters depends on the satellite's design. Let $G$ be the set of ground stations, and $R_g$ be the number of receivers at ground station $g$ for each $g \in G$. Let $F \subset G \times G$ be the set of ground station pairs that can be and desire to be connected, and let $L_j$ be the maximum number of connections ground station pair $j \in F$ will accept between them. We assume $L_j \le \min\{R_{g_l},R_{g_m}\}$ for pair $j = (g_l, g_m).$
\subsection{Dual Downlink Architecture}\label{dd_architecture}
In this work, we consider a dual downlink architecture as shown in Figure \ref{fig:dual_downlink}. Here, entangled pairs of qubits are generated by satellites and transmitted to two receiving stations on the ground via free space. We assume satellites have no onboard quantum memories, so the transmission of each qubit in an entangled pair must occur immediately. This architecture, while pragmatic for its limited use of quantum memory, is limited in that only pairs of ground stations that are in view of the same satellite can be connected. Therefore connections are limited by geographic distance\footnote{In our simulations, which use low Earth orbit satellites and require that satellites be a minimum of 20 degrees above the horizon, the average satellite can directly connect ground stations at most 2,400 kilometers apart.}, depending upon satellite location. This limitation can be overcome with inter-satellite links which is out of the scope of this work. 
In this work we focus on the problem of maximizing entanglement distribution rate under the constraint that there are no inter-satellite links. This is a useful metric in its own right, and serves as a preliminary metric for evaluating the distribution rate of protocols using entanglement swapping and inter-satellite links.

\subsection{Entanglement Generation Source}\label{source}
In our setup, bipartite entanglements are generated using a Spontaneous Parametric Down-Conversion (SPDC) based dual-rail polarization entangled pair source located on the satellite \cite{panigrahy2022optimal,dhara2022heralded}. We assume the generated entanglements have the following form.
\begin{align}
		\ket{\psi^\pm}&\!=   N_0 \bigg[\!\sqrt{p(0)} \ket{0,0;0,0}\!\nonumber\\
        &+\! \sqrt{\frac{p(1)}{2}} \left(\ket{1,0;0,1}\pm\ket{0,1;1,0} \right) \nonumber\\
		&+  \sqrt{\frac{p(2)}{3}}\left(\ket{2,0;0,2}\pm \ket{1,1;1,1}+\ket{0,2;2,0}\right)\bigg],
		\label{eqn:srcnative}
\end{align}
where $N_0$ is a normalization constant and 
\begin{equation}
p(n) = (n+1)\frac{{N_s}^n}{(N_s + 1)^{n+2}}. 
\label{eqn:geometric_dist}
\end{equation}
Here, $N_s$ denotes the mean photon number per mode, also known as the pump power. The source outputs a superposition state comprised of the following three elements: (i) an entangled pair ($[\ket{1,0;0,1}\pm\ket{0,1;1,0}]/\sqrt{2}$), (ii) the vacuum ($\ket{0,0;0,0}$), and (iii) the unwanted two-pair photon terms ($[\ket{2,0;0,2}\pm \ket{1,1;1,1}+\ket{0,2;2,0}]/\sqrt{3}$). Note that, choosing the right amount of pump power is important as   higher pump power not only enhances the rate of entanglement production but also increases the rate of noisy two photon pair emissions.

\subsection{Loss and Noise}\label{loss_noise}
The entanglements generated at the source are split and transmitted to their respective ground stations through free space. While undergoing transmission, entanglements experience different forms of losses, with two significant ones being -  free space loss and atmospheric loss. Free space loss primarily occurs due to diffraction and scales quadratically, whereas  absorption and scattering contribute to atmospheric loss, which grows as an exponential function of distance. Unlike ground based transmissions where losses are exponential, the majority of the losses in free space are the lower quadratic losses. 

In this work. we consider two sources of noise - (i) Two pair photon emissions, and (ii) Background photon flux. Due to two pair photon emissions, losses in the channel can result in the loss of a single photon from two pair photons, leading them to resemble entangled pairs. As these pairs are not inherently entangled, they cause a decrease in fidelity of a distributed pair. As stated earlier, two photon pair emissions can be controlled by choosing appropriate pump power. 

Background photons are being continuously generated from different light sources (e.g. sun, moon, stars). They  introduce noise, since it is difficult to distinguish random and uncorrelated background photons from source-generated entangled photons. We model the noise introduced by background photons as an increase in the number of dark click counts at the detector at the receiver. We treat the arrival of any  background photon as a spurious detection event registered by the receiver even though no logical photon is present in the system. Typically, the background photon flux is orders of magnitude higher during daytime compared to night time.

\subsection{Scheduling in a time snapshot}
In this work, we assume that time is divided into slots, and the scheduling problem is solved for each slot. This means that the satellite-to-ground station pair-assignments, which are solutions to the scheduling problems presented in this paper, are intended to be used only for a brief period. In the next timeslot, the pair-assignments must be recalculated with new inputs. This assumption simplifies the problem, and reflects likely network deployment limitations in a few ways. In practice, the changes to the input parameters of the problem (ambient conditions, such as weather) may occur at arbitrary times and may involve only a small set of satellites and ground stations. In this work, we do not model the continuous time version of the problem and instead treat it as a discrete time system where scheduling changes occur at the beginning of each time unit (slot). 

Also, there may be a cost associated with such changes (e.g. delay due to realignment of satellite transmitters or ground station receivers), which might influence scheduling decisions. Any scheduling algorithm in deployment should attempt to reduce the frequency with which communications equipment will need to be aligned, a consideration that requires coordination across multiple slots in time, something we do not address in this work. However, it is worth noting that emerging entanglement generation sources, like the \emph{zero-added-loss multiplexing} (ZALM) sources \cite{Chen23}, can make the changes with very low overhead. Designing scheduling policies that account for switching cost is a topic of our future work.

% In practice, this means that the satellite-to-ground station pair-assignments, which are solutions to the scheduling problems presented in this paper, are intended to be used only for a brief period. Then, when the input parameters are no longer sufficiently reflective of real-world conditions (a metric outside the scope of this work), the pair-assignments must be recalculated with new inputs. This assumption simplifies the problem, and reflects likely network deployment limitations in a few ways. As discussed in Section \ref{dd_architecture}, entangled pairs of qubits must be transmitted simultaneously, so we can gain no advantage by considering where satellites are moving. Furthermore, crucial ambient conditions such as weather can change rapidly and unpredictably, making planning ahead of limited value.

% Independently scheduling each time snapshot has some limitations, however. We ignore the time and energy it takes to reposition satellite transmitters or ground station receivers, and instead assume that all ground stations and satellites can be immediately and perfectly aligned. Any scheduling algorithm in deployment should also attempt to reduce the frequency with which communications equipment will need to be aligned, a consideration that requires coordination across multiple slots in time, something we do not address in this work. Additionally, in this work we ignore classical communication delays, which could be restrictive when we consider non-local algorithms that require up-to-date information on local conditions.

\section{Quantum Satellite Scheduling Problem}
We now formally introduce the Quantum Satellite Scheduling Problem (QSSP).

For each $i \in S$ and $j \in F$, let the decision variables $x_{ij}(t)$ represent the number of connections to be made between ground stations $g_1$ and $g_2$ via satellite $i$ at time slot $t$, for $j=(g_1, g_2)$. As the number of connections is discrete and non-negative, $x_{ij}(t) \in \Znon$ for all $i \in S,\;j \in F$. Let $w_{ij}(t)$ be the acquired reward of connecting the ground station pairs in $j$ using satellite $i$ at time $t$. This can be variously defined, but in our simulations (described in section \ref{simulations}) we use the entanglement distribution rate as the weight associated with a connection, which is calculated with respect to atmospheric and free space distance between satellite $i$ and each ground station in $j$. Note that if either ground station in $j$ is out of range for satellite $i$, $w_{ij}(t) = 0$.

In its most general form, the problem of scheduling transmissions from satellites to ground station pairs with arbitrary resource/utility limits on satellites, ground station pairs, and individual ground stations, can be formalized as:\\

%\begin{prob}[QSSP]
%\label{matching}
\noindent{\bf{QSSP \cite{panigrahy2022optimal}:}} Given $S, F,$ and $G$, and $T_i, L_j, R_g,$ and $w_{ij}(t)$ for all $i \in S, j \in F, g \in G$ (i.e. values for the elements of Table \ref{table:notation}), find values for $x_{ij}(t)$ at each time slot $t$ that 
\begin{align}
\text{maximize} &\sum_{i\in S, j\in F} w_{ij}(t)x_{ij}(t)\\
\text{subject to}\nonumber\\
    \forall i \in S:& \sum_{j \in F} x_{ij}(t) \le T_i,\\
    \forall j \in F:& \sum_{i \in S} x_{ij}(t) \le L_j,\\
    \forall g \in G:& \sum_{i \in S}\sum_{j \in F|g \in j} x_{ij}(t) \le R_g, \label{receivers_constraint}\\
    \forall i \in S, \forall j \in F:& \quad x_{ij}(t) \in \{0,1,\cdots\}.
\end{align}
%\end{prob}

% \begin{figure}
%     \centering
%     \includegraphics[width=0.4\textwidth]{figs/3-dimensional-matching.svg}
%     \caption{3D-matching}
%     \label{fig:tripartite_matching}
% \end{figure}

\subsection{NP-hardness of QSSP}
The main result of this Section is as follows.

\begin{thm}
\label{marching_is_hard}
The Quantum Satellite Scheduling Problem is NP-hard.
\end{thm}

To show that QSSP is NP-hard, we give a reduction from  a known NP-hard problem: 3-Dimensional Matching.\\

\noindent{\bf{3D Matching \cite{Karp1972}:}} Let $G^{3D}=\left(V_1\cup V_2 \cup V_3, E\right)$ be an undirected tripartite hypergraph such that $E \subseteq V_1 \times V_2 \times V_3$. Find the maximum cardinality set of hyperedges $E^* \subseteq E$ such that $(v_1, v_2, v_3) \ne (v_1', v_2', v_3')$ and $(v_1, v_2, v_3), (v_1', v_2', v_3') \in E^* \implies v_1 \ne v_1', v_2 \ne v_2',$ and $v_3 \ne v_3'$.\\

For any arbitrary instance $G^{3D}\left(V_1\cup V_2 \cup V_3, E\right)$ of the 3D Matching problem, we build an instance of QSSP\footnote{We remove the time dependency, i.e., use $w_{ij}, x_{ij}$ instead of $w_{ij}(t), x_{ij}(t)$ for brevity.}, i.e., we construct an input:
\begin{align*}
    &S, G, F, \left\{T_i: i \in S\right\}, \left\{R_g: g \in G\right\}, \left\{L_j: j \in F\right\},\\
    &\left\{w_{ij}:i \in S, j \in F\right\}
\end{align*}
to QSSP. Initially, let $S = F = G = \phi.$ For every hyperedge $(v_k, v_l, v_m) \in E$, update $S\leftarrow S\cup\{i_k\}$, $F\leftarrow F\cup\{j=(g_l, g_m)\}$, and $G\leftarrow G\cup\{g_l, g_m\}$. Lastly, let $w_{ij} = T_i = L_j = R_g = 1$ for all $i \in S, j \in F, g \in G$.

Solving QSSP on this input gives a solution $x_{ij} \in \{0,1\}$ ($x_{ij} \le 1$ since $\forall i \in S: T_i = 1$, so $\forall i \in S, j \in F: x_{ij} \le \sum_{j' \in F}x_{ij'} \le 1$). Let $E'$ be the set of hyperedges such that $(v_k, v_l, v_m) \in E'$ if and only if  $x_{i_k(g_l, g_m)} = 1$. We aim to show that $E'$ is a solution to the 3D Matching problem on input $V_1, V_2, V_3, E$.

First we show that $E'$ is a matching of $G^{3D}$. Note that, because $T_i = L_j = R_g = 1$ for all $i \in S, j \in F, g \in G$ and by construction, we have the following.
\begin{align*}
    &(v_k, v_l, v_m) \ne (v_k', v_l', v_m') \text{ and } \\&(v_k, v_l, v_m), (v_k', v_l', v_m') \in E'\\
    \implies &(i_k,j=(g_l, g_m)) \ne (i_k', j' = (g_l', g_m'))\text{ and }\\&x_{i_kj} = x_{i_k'j'} = 1\\
    \implies &i_k \ne i_k', g_l \ne g_l', \text{ and } g_m \ne g_m'\\
    \implies &v_k \ne v_k', v_l \ne v_l', \text{ and } v_m \ne v_m'
    % &\implies
    % &(i_k,j=(g_l, g_m)) \ne (i_k', j' = (g_l', g_m'))\\
    % &\text{and } x_{i_kj} = x_{i_k'j'} = 1 \\
    % \implies&\\
    % &i_k \ne i_k', g_l \ne g_l', \text{ and } g_m \ne g_m'
\end{align*}
Now it remains to show that $E'$ is a maximum cardinality matching.

Suppose that there exists some $E^* \subseteq E$ such that $(v_k, v_l, v_m) \ne (v_k', v_l', v_m')$ and $(v_k, v_l, v_m), (v_k', v_l', v_m') \in E^* \implies v_k \ne v_k', v_l \ne v_l',$ and $v_m \ne v_m'$, and $|E^*| > |E'|$. Let $x^*_{i_kj}$ be defined as $x^*_{i_kj} = 1$ when $(v_k, v_l, v_m) \in E^*$, and $x^*_{i_kj} = 0$ when $(v_k, v_l, v_m) \in E\setminus E^*$ with $j=(g_l, g_m)$. Also, $\forall i_k \in S,\;j \in F, \text{ and } g \in G,$ we have,
\begin{align*}
&\sum_{j' \in F} x^*_{i_kj'}
    = |\{j'=(g_l,g_m)|(v_k, v_l, v_m)\in E^*\}| \le 1,\\
&\sum_{i'_k \in S} x^*_{i'_kj}(t)
    = |\{i'_k|(v_k, v_l, v_m)\in E^*\}| \le 1,\\
&\sum_{i'_k \in S}\sum_{j' \in F|g \in j'} x^*_{i'_kj'}\\
    &= |\{i'_k, j'=(g_l,g_m)|(v_k, v_l, v_m)\in E^*\}| \le 1.
\end{align*}
Thus the vector: $[x^*_{ij}, \forall i\in S, j \in F]$ satisfies the constraints defined for the above instance of QSSP. But, we also have,
\begin{equation*}
    \sum_{i\in S,j \in F}x^*_{ij} = |E^*| > |E'| = \sum_{i\in S,j \in F}x_{ij},
\end{equation*}
which is a contradiction, since $[x_{ij}, \forall i\in S, j \in F]$ is the optimal solution of the defined instance of QSSP. Therefore, $\nexists E^*: |E^*| > |E'|$ and $E'$ is a solution to the 3D Matching problem. Because any arbitrary instance of the 3D Matching problem can be transformed into an instance of QSSP in polynomial time, and 3D Matching is NP-hard, QSSP is also NP-hard.

\begin{remark}
    It's important to highlight that QSSP is more difficult than its classical counterpart. The difficulty of this problem arises from the concept of assigning satellites to ground station pairs, as opposed to individual ground stations, while still facing resource constraints on individual ground stations. Pair assignment, an uncommon feature in classical satellite networks, adds complexity in the context of quantum satellite networks.
\end{remark}

\subsection{Special Cases}
While the most general version of QSSP is NP-hard, below we present a few special cases for which polynomial time solutions exist. 
\subsubsection{Satellites outnumbering ground stations}
We assume a scenario where the satellites significantly outnumber the ground stations. This scenario ensures the availability of a satellite for connecting any two ground stations within each other's range. Under this assumption, we can solve the problem of pairing ground stations together before assigning satellites to them. Therefore, removing the satellite transmitter constraints, we get the following optimization problem.
\begin{align}\label{eq:bmatching}
    \text{maximize} \sum_{j\in F} w_jx_j\\
    \text{subject to: } \forall j \in F:&\quad x_j \le L_j,\nonumber\\
    \forall g \in G:& \sum_{j \in F|g \in j} x_j \le R_g,\nonumber\\
    \forall j \in F:& \quad x_{j} \in \{0,1,\cdots\}.\nonumber
\end{align}
\noindent Here, the weights $w_j$, represent increasing affine functions of the distance between ground stations in pair $j.$ For this scenario, we can construct a graph $U(V,E)$ where $V = G$ and $E=F$. We set $w_j$ and $L_j$ to be the weight and capacity of edge $j\in E,$ respectively. Define $b(g) = R_g$ as the capacity of vertex $g \in V.$ Then solving problem \eqref{eq:bmatching} is equivalent to solving the weighted capacitated b-matching problem for $U$. The separation algorithm by Letchford et al. \cite{Letchford04} can be used to solve Problem \eqref{eq:bmatching} with a running time complexity of $O(|G|^2|F|\log(|G|^2/|F|)).$ 
% \end{prob}

% Problem \ref{gs_matching} is equivalent to the problem of finding a maximal b-Matching in an arbitrary graph with weighted and capacitated edges, defined as:

% \begin{prob}
% \label{b-matching}
% Given a graph $G = (V, E)$, a capacity $b_i$ for each vertex $i \in V$, and a weight $w_e$ and a capacity $c_e$ for each edge $e \in E$, find values for $x_e \in \mathbb{Z}^{\ge 0}$ that together maximize
% \begin{equation*}
%     \sum_{e \in M}w_ex_e
% \end{equation*}
% such that
% \begin{align*}
%     \forall e \in E:& x_e \le c_e\\
%     \forall i \in V:& \sum_{e \in M: i \in e} x_e \le b_i
% \end{align*}
% \end{prob}

\subsubsection{No resource constraints at ground stations}
% % \begin{figure}
% %     \centering
% %     \includegraphics[width=0.4\textwidth]{figs/bipartite sat matching.png}
% %     \caption{Problem \ref{sim_matching_prob} is weighted bipartite matching}
% %     \label{fig:bipartite_matching}
% % \end{figure}
Such a scenario corresponds to removing the constraints on the number of receivers (Constraint \eqref{receivers_constraint} in QSSP) at each ground station, or, equivalently, setting $R_g = \infty$ for all $g \in G$. This can be accomplished by including many receivers at each ground station which makes QSSP tractable. For this case, we can construct a bipartite graph: $U(V, E)$ with bi-partitions $S$ and $F$, such that $V = S \cup F$ and $E = C = \{(i,j), \forall i\in S, \forall j\in F, \text{ and } w_{ij} > 0\}.$ We set $w_{ij}$ as the weight of edge $(i,j)\in E.$ Solving QSSP, when $R_g = \infty,$ is analogous to solving the uncapacitated  weighted b-matching problem for $U.$ QSSP, in this case, can be solved using the modified minimum cost flow algorithm by Anstee et al. \cite{Anstee87} with a running time complexity of $O((|S|+|F|)^2|C|).$ This scenario ($R_g = \infty$) suggests a heuristic that we present in the next subsection.
%We define $b(i) = T_i, \forall i \in S$ and $b(j) = L_j, \forall j \in F$ as capacities of vertices.
% In our simulations, we consider a further simplified version of this problem in which $T_i = 1$ for all $i \in S$ and $L_j = 1$ for all $j \in F$:

% \begin{prob}\label{sim_matching_prob}
%     Given values for entries in Table \ref{table:notation}, for all $i \in S, j \in F$ such that $w_{ij} > 0$, find values for $x_{ij} \in {0,1}$ that together maximize
%     \begin{equation*}
%         \sum_{i \in S, j \in F}w_{ij}x_{ij}
%     \end{equation*}
%     subject to
%     \begin{align*}
%         \forall i \in S:& \sum_{j \in F}x_{ij} \in \{0, 1\}\\
%         \forall j \in F:& \sum_{i \in S}x_{ij} \in \{0, 1\}
%     \end{align*}
% \end{prob}

% This problem is equivalent to weighted bipartite matching, which can be solved optimally in polynomial time. We use the solution in \cite{galil1986efficient} in our simulations as a comparison to alternative heuristic algorithms which can be extended to broader definitions of the problem.
% entanglement distribution rate of we test are:
\begin{algorithm}
\caption{Heuristic Algorithms for QSSP}\label{algos}
\begin{algorithmic}[1]
\Procedure{UpdateState}{$i, j, C$}
\State $x_{ij} \gets x_{ij} + 1$
\State $(g_l, g_m) = j$
\State $T_i \gets T_i - 1$; $L_j \gets L_j - 1$
\State $R_{g_l} \gets R_{g_l} - 1$; $R_{g_m} \gets R_{g_m} - 1$
\If {$T_i = 0$} $C \gets C\setminus\{(i,j'),\forall j'\in F \}$
\EndIf
\If {$L_j = 0$} $C \gets C\setminus\{(i',j),\forall i'\in S \}$
\EndIf
\If {$R_{g_l} = 0$} 
\State $C \gets C\setminus\{(i',j'), \forall i'\in S, \forall j'|g_l\in j'\}$
\EndIf
\If {$R_{g_m} = 0$}
\State $C \gets C\setminus\{(i',j'), \forall i'\in S, \forall j'|g_m\in j'\}$
\EndIf
\EndProcedure
\Procedure{RANDOM}{C}
\While {$C\ne\phi$} 
\State Pick $(i,j)$ uniformly at random from $C$.
\State Call Procedure UpdateState($i, j, C$).
\EndWhile
\EndProcedure
\Procedure{LOCAL\_GREEDY}{C}
\While {$C\ne\phi$} 
\State $F' = F \cap \{j'|(i',j')\in C, \forall i'\in S\}$
\State Pick $j$ uniformly at random from $F'$.
\State Pick $i\in S$ s.t. $(i,j)\in C$, $i$ maximizes $w_{ij}$.
\State Call Procedure UpdateState($i, j, C$).
\EndWhile
\EndProcedure
\Procedure{GLOBAL\_GREEDY}{C}
\While {$C\ne\phi$} 
\State Pick $(i,j)$ from $C$ that maximizes $w_{ij}$.
\State Call Procedure UpdateState($i, j, C$).
\EndWhile
\EndProcedure
\State Initialize $C = \{(i,j), \forall i\in S, \forall j\in F, \text{ and } w_{ij} > 0\}$, $x_{ij} = 0, \forall i \in S, j \in F$. \label{init}
\end{algorithmic}
\end{algorithm}
\subsection{Heuristic Algorithms}
Since the most general version of QSSP is NP-hard and lacks a polynomial time solution, we shift our attention to developing heuristic algorithms. We consider three heuristic algorithms corresponding to different orderings of the connections - (i) RANDOM, (ii) LOCAL\_GREEDY, and (iii) GLOBAL\_GREEDY. The techniques are presented as procedures in Algorithm \ref{algos}. 

Each technique starts with an instance of all available connections $C$ (See line \ref{init} in Algorithm \ref{algos}). Subsequently, a specific connection is selected based on various criteria, each corresponding to a distinct heuristic rule. For example, in the RANDOM algorithm, a connection is chosen uniformly at random from $C$. In LOCAL\_GREEDY, a ground station pair is first chosen uniformly at random, and then a satellite is selected to establish the connection in a manner that maximizes the entanglement distribution rate of the pair. In GLOBAL\_GREEDY, the connection that maximizes the entanglement distribution rate among all connections in $C$ is selected. Upon selecting a connection, the corresponding resources utilized to serve this connection are removed from the available resource pool for all heuristic methods. If a satellite or a ground station depletes all its resources, all of its connections are removed from $C$. This procedure continues until $C$ becomes empty.
\begin{algorithm}
\caption{GREEDY\_BACKOFF algorithm for QSSP}\label{algo-backoff}
\begin{algorithmic}[1]
\Procedure{GREEDY\_BACKOFF}{C}
\State $\mathtt{done \gets False}$\label{init}
\While{$\mathtt{done}$ is $\mathtt{False}$}
\State Use modified minimum cost flow algorithm by Anstee et al \cite{Anstee87} to find an uncapacitated  maximum weight b-matching with edge weights $w_{ij}$ to assign $x_{ij}$ for all $i \in S, j = (g_l, g_m) \in F$ such that $i, j$ are not already at capacity for $T_i$, $L_j$, $R_{g_l}$, and $R_{g_m}$.
%\State Use Edmonds blossom algorithm \cite{galil1986efficient} with edge weights $w_{ij}$ to assign $x_{ij}$ for all $i \in S, j = (g_l, g_m) \in F$ such that $i, j$ are not already at capacity for $T_i$, $L_j$, $R_{g_l}$, and $R_{g_m}$.
\State $\mathtt{done \gets True}$
\State $\mathtt{compliant \gets False}$
\While{$\mathtt{compliant}$ is $\mathtt{False}$}
\State $\mathtt{compliant \gets True}$
\If{any constraints are violated}
\State $\mathtt{done \gets False, compliant \gets False}$
\State Remove the non-compliant connection with the lowest $w_{ij}$ value.
\EndIf
\EndWhile
\EndWhile
\EndProcedure
\State Initialize $C = \{(i,j), \forall i\in S, \forall j\in F, \text{ and } w_{ij} > 0\}$, $x_{ij} = 0, \forall i \in S, j \in F$.
\end{algorithmic}
\end{algorithm}

Additionally, we introduce a fourth heuristic method, GREEDY\_BACKOFF, which relies on an optimal algorithm for uncapacitated  maximum weight b-matching (Anstee et al algorithm \cite{Anstee87}) between satellites and ground station pairs. This is outlined in Algorithm \ref{algo-backoff}. The algorithm involves initially establishing a b-matching between satellites and ground stations, with edge weights corresponding to entanglement distribution rate of the connections. Any violation of the ground station receiver constraint (Constraint \eqref{receivers_constraint} in QSSP) is addressed by removing connections with the lowest entanglement distribution rate. The process is repeated until all resources are exhausted, and no more connections remain to be served. Note that GREEDY\_BACKOFF is optimal when $R_g = \infty, \forall g \in G.$ 

\section{Performance Evaluation}\label{simulations}
% \begin{figure}
%     \centering
%     \includegraphics[width=0.4\textwidth]{figs/starlink_placements.png}
%     \caption{3,967 low earth orbit satellites in the Starlink constellation}
%     \label{fig:starlink_placements}
% \end{figure}
\begin{figure*}[htbp]
    \centering
    \begin{subfigure}[t]{0.32\textwidth}\includegraphics[width=\textwidth]{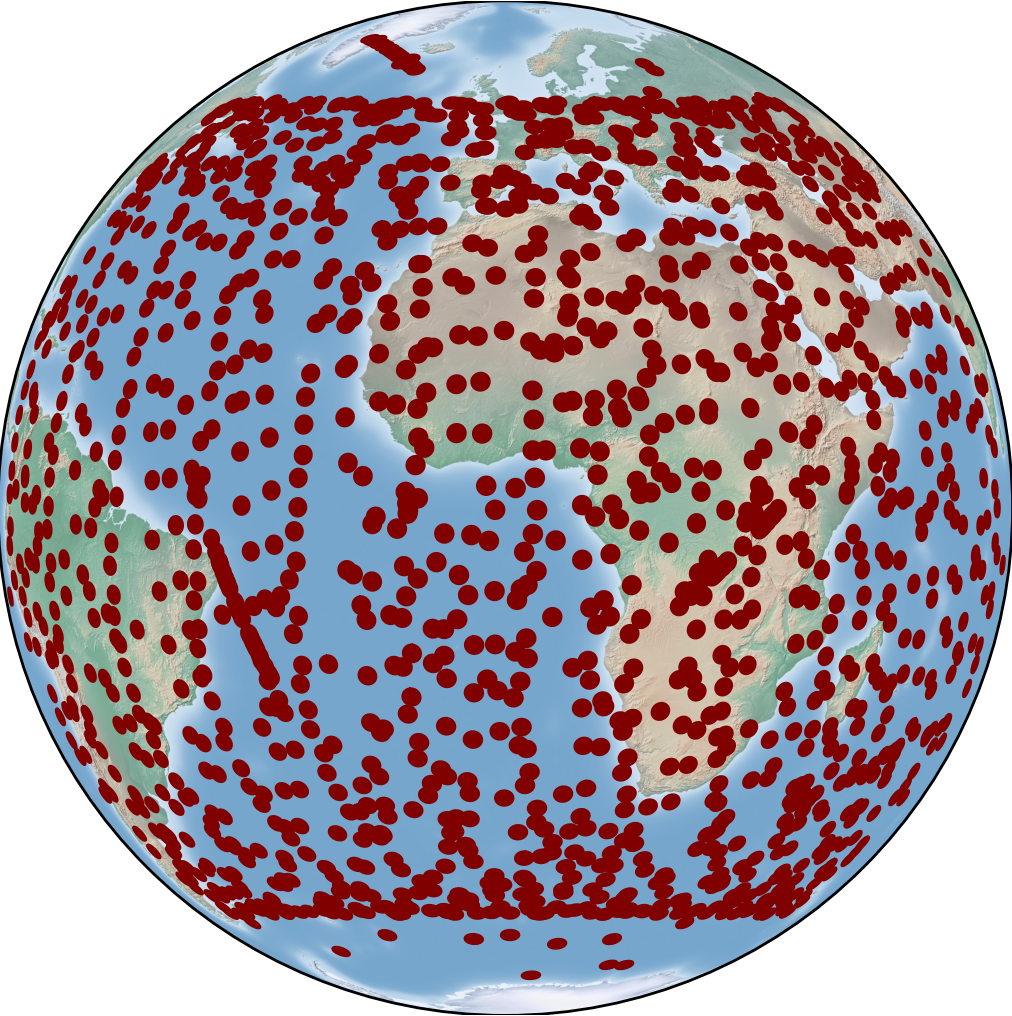}
    \caption{}
    \end{subfigure}
    \begin{subfigure}[t]{0.32\textwidth}\includegraphics[width=\textwidth]{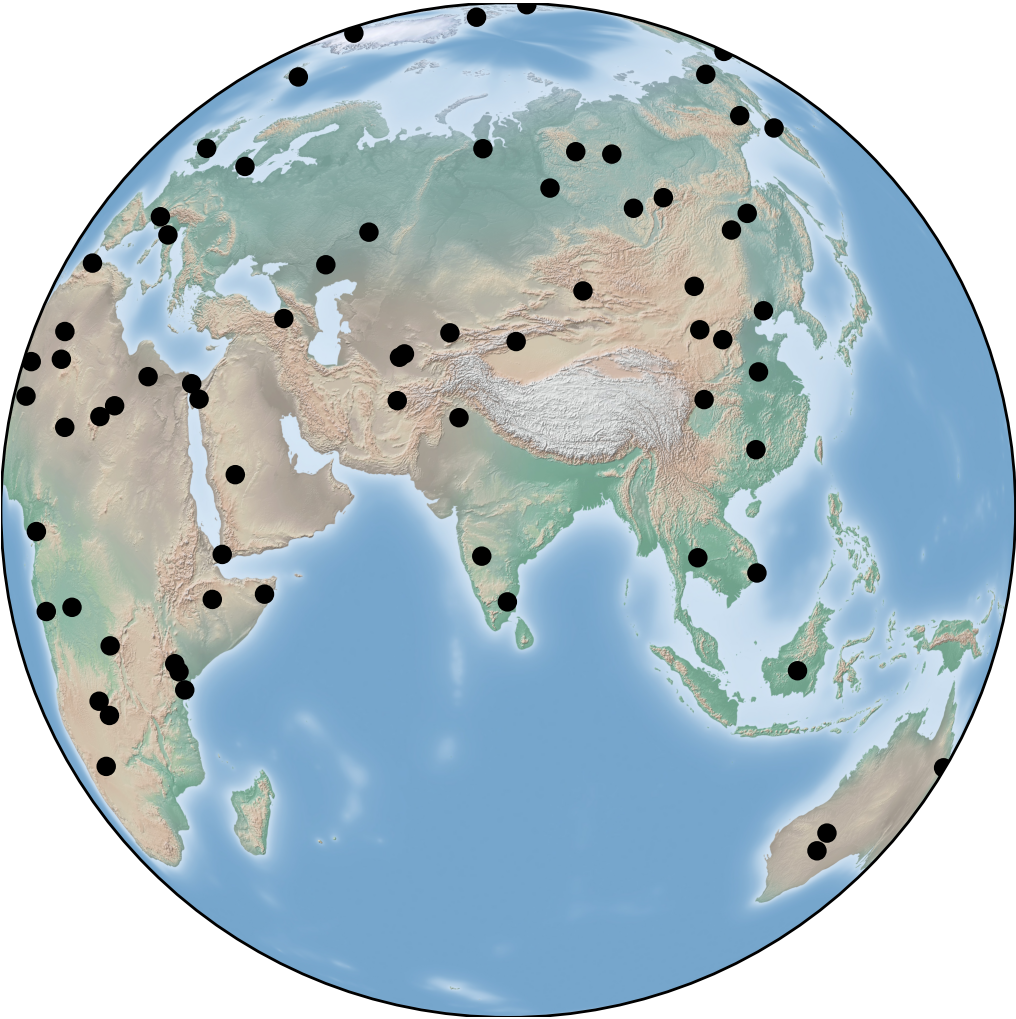}
    \caption{}\end{subfigure}
    \begin{subfigure}[t]{0.32\textwidth}\includegraphics[width=\textwidth]{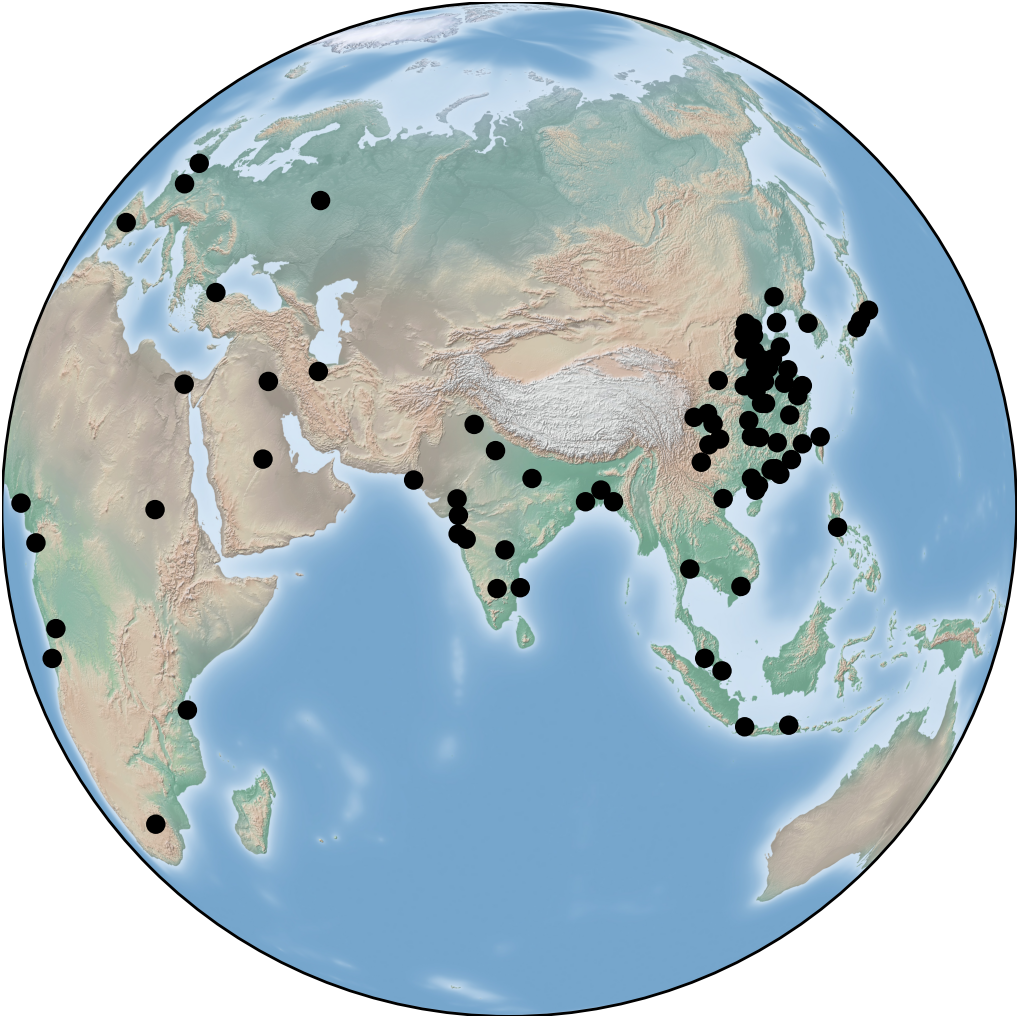}\caption{}\end{subfigure}
    \caption{Satellites and ground station distributions, $|S| = 3967, |G| = 100.$ (a) Satellites in the Starlink LEO mega-constellation. (b) Ground stations are placed uniformly at random on the land. (c) Ground stations are chosen based on highest population density.}
    \label{fig:sat_gs_distributions}
\end{figure*}

In this section, we evaluate the solutions generated by various heuristics and investigate the impact of factors such as satellite and ground station density and ground station placement on their performance. We test the performance of these heuristics using the Starlink LEO mega-constellation as a case study. The Starlink constellation is a leading candidate among all mega-constellations intended for classical satellite internet deployment. It yields very large scale instances of QSSP, and the resulting instances cannot achieve optimal solutions due to the computational hardness of QSSP. We detail the Starlink case in the following section.

\begin{figure*}[htbp]
    \centering
    \begin{subfigure}[t]{0.32\textwidth}\includegraphics[width=\textwidth]{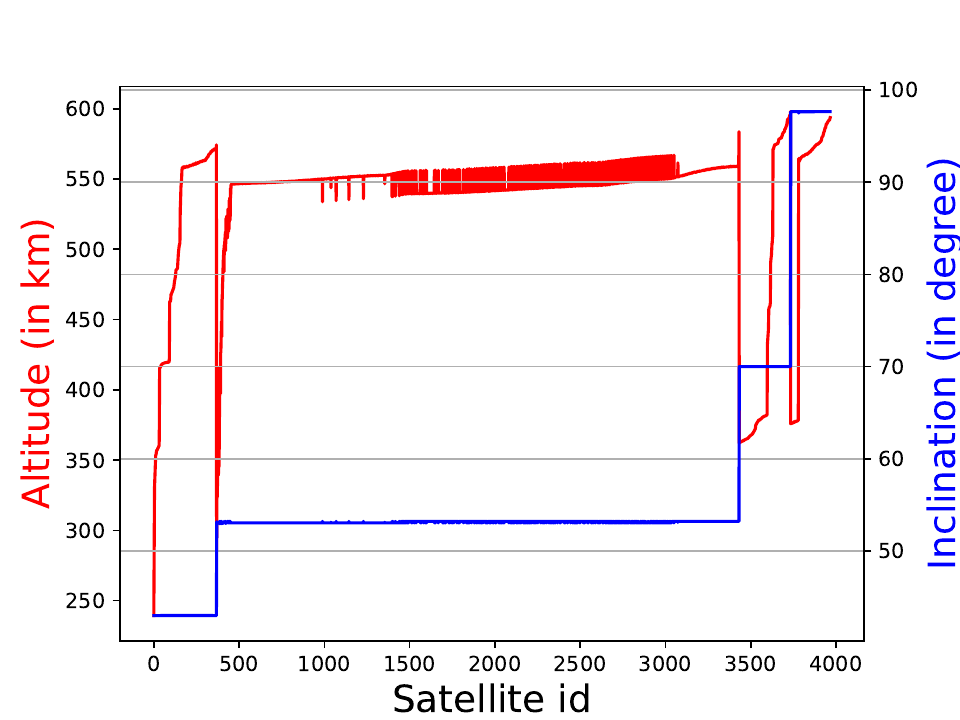}
    \caption{}
    \end{subfigure}
    \begin{subfigure}[t]{0.32\textwidth}\includegraphics[width=\textwidth]{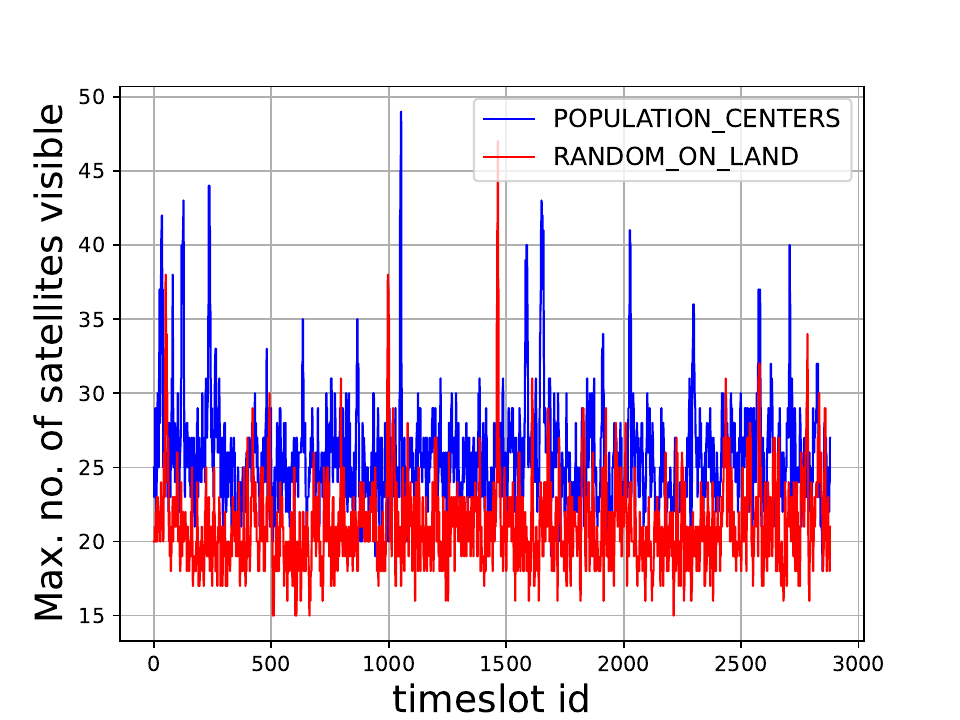}
    \caption{}\end{subfigure}
    \begin{subfigure}[t]{0.32\textwidth}\includegraphics[width=\textwidth]{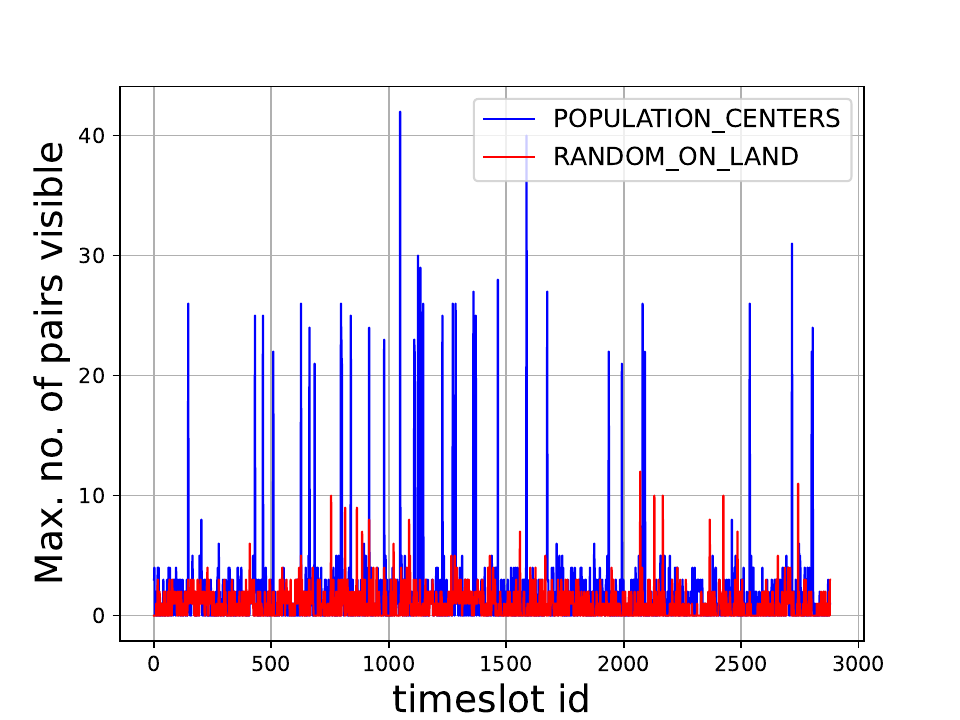}\caption{}\end{subfigure}
    \begin{subfigure}[t]{0.32\textwidth}\includegraphics[width=\textwidth]{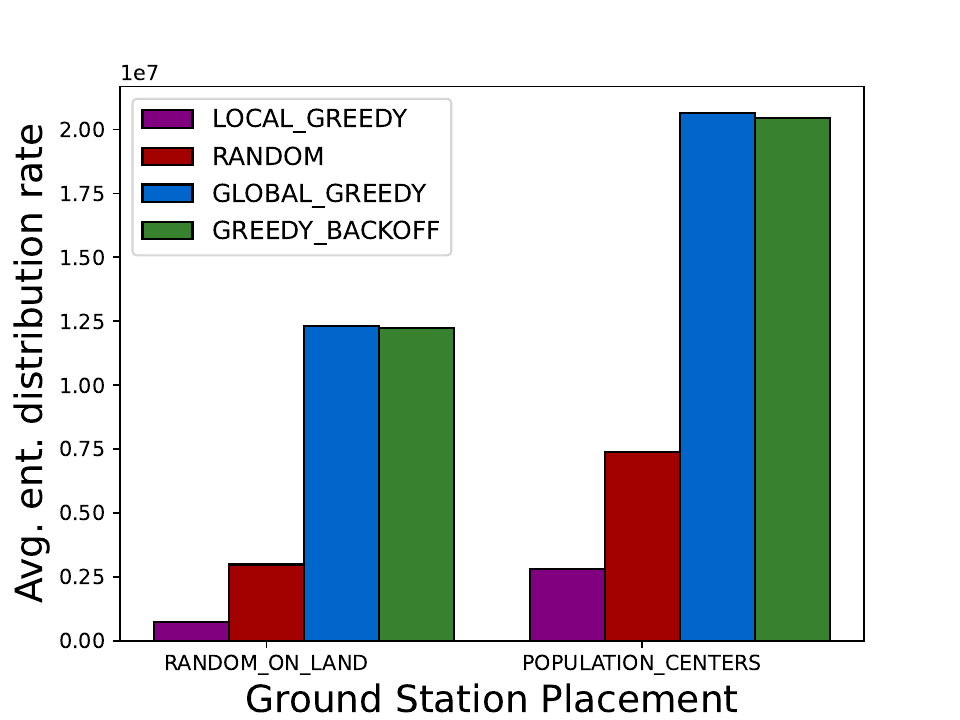}
    \caption{}
    \end{subfigure}
    \begin{subfigure}[t]{0.32\textwidth}\includegraphics[width=\textwidth]{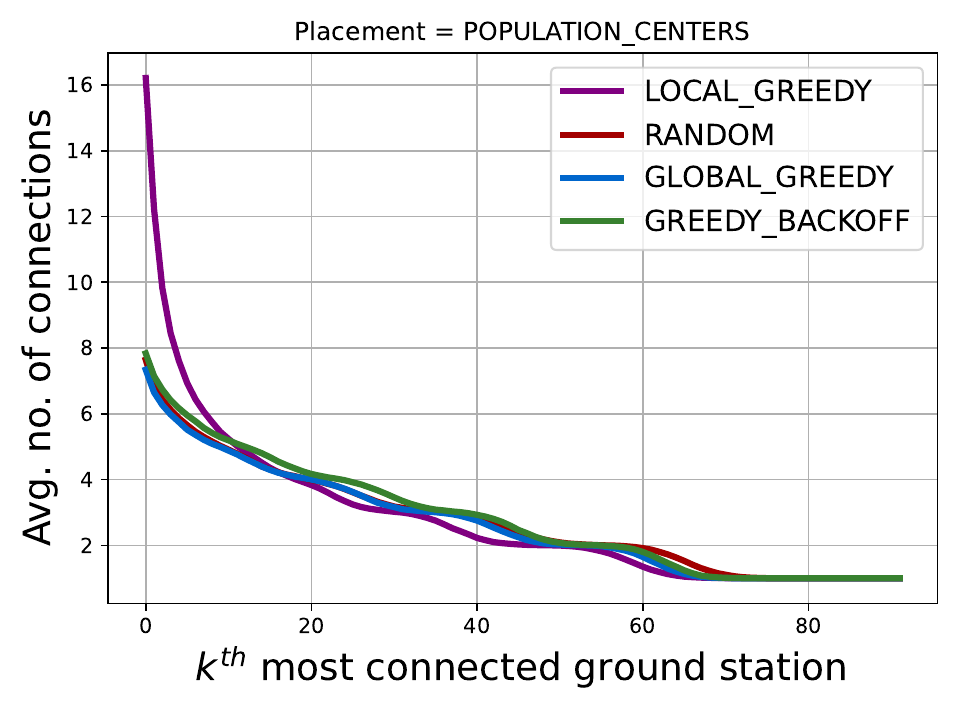}\caption{}\end{subfigure}
    \begin{subfigure}[t]{0.32\textwidth}\includegraphics[width=\textwidth]{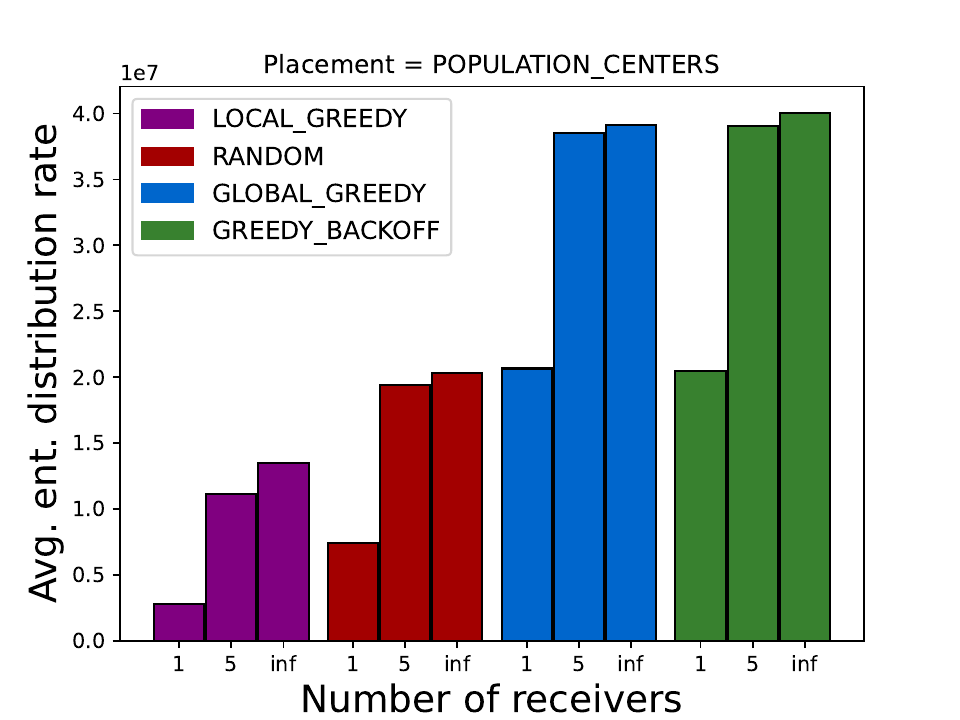}\caption{}\end{subfigure}
    \begin{subfigure}[t]{0.32\textwidth}\includegraphics[width=\textwidth]{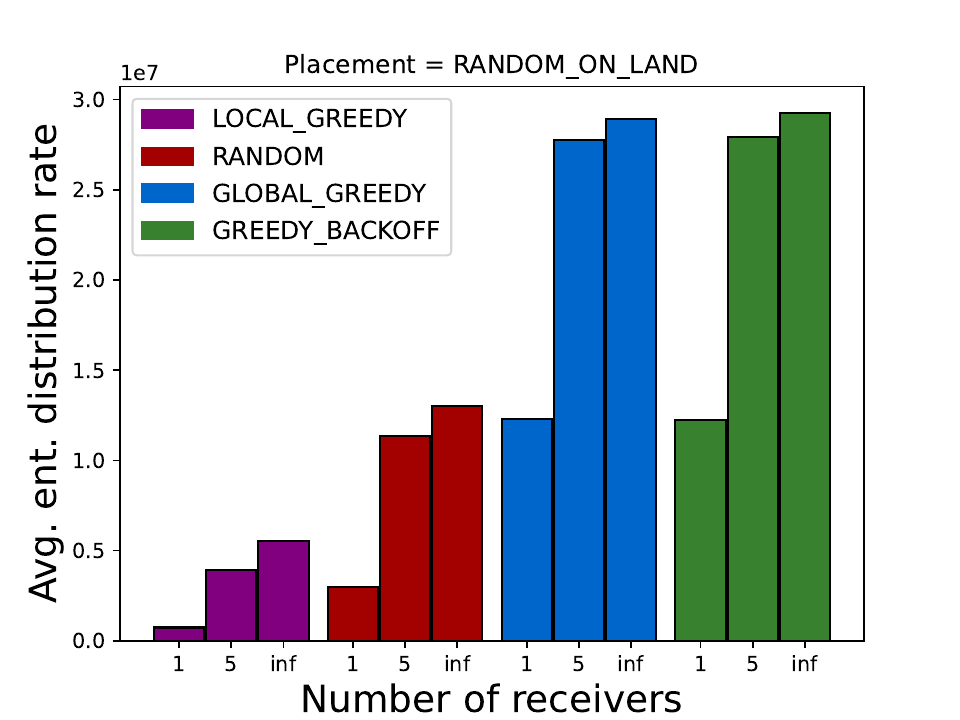}
    \caption{}\end{subfigure}
    \begin{subfigure}[t]{0.32\textwidth}\includegraphics[width=\textwidth]{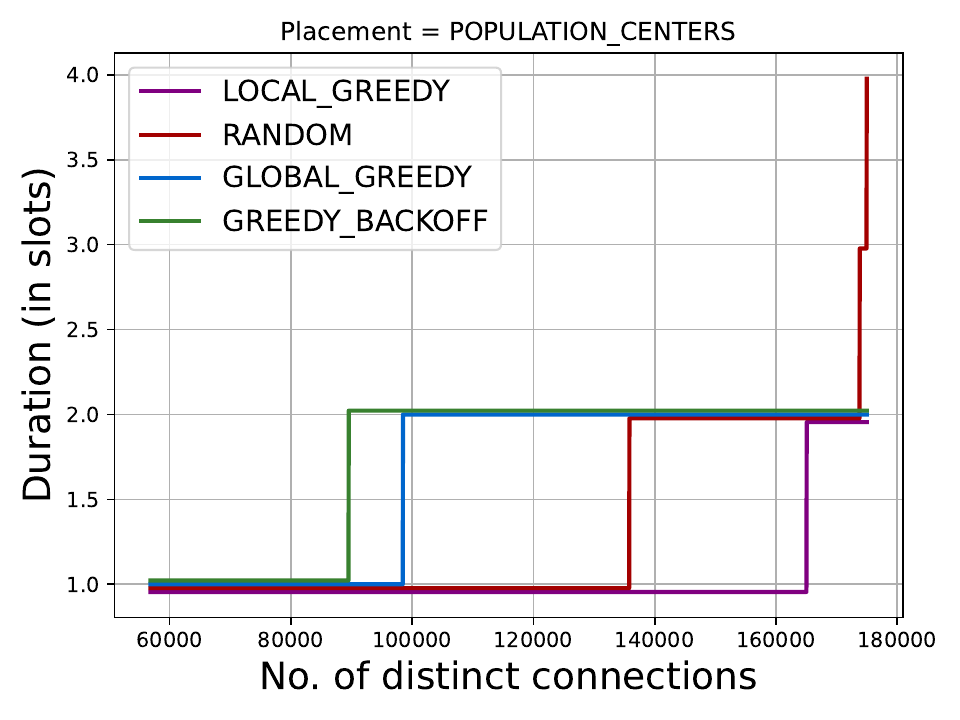}
    \caption{}\end{subfigure}
    \begin{subfigure}[t]{0.32\textwidth}\includegraphics[width=\textwidth]{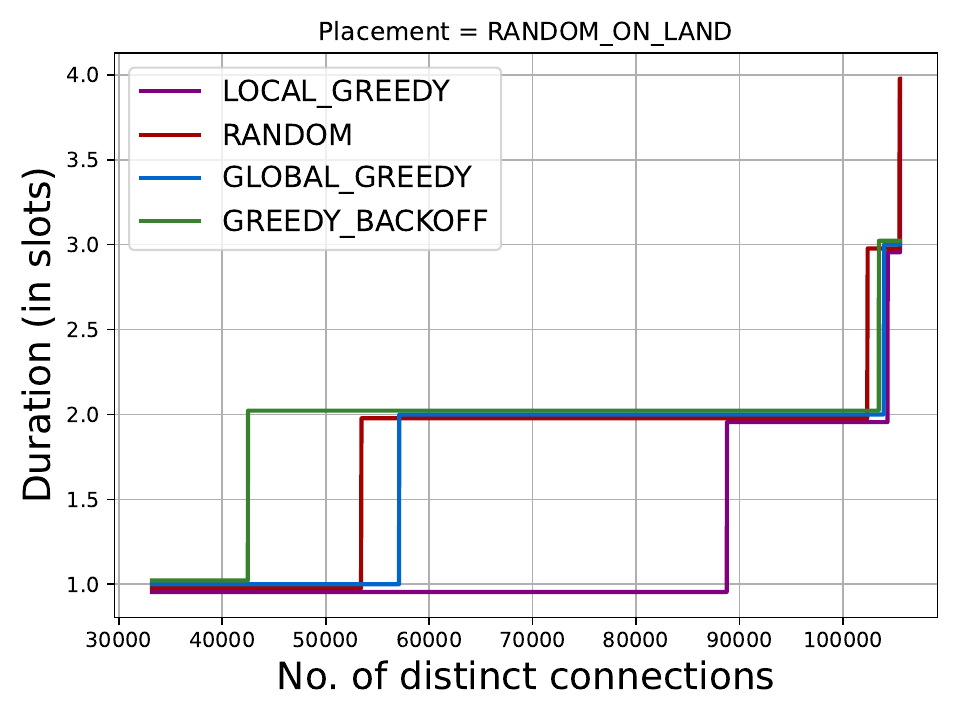}
    \caption{}
    \end{subfigure}
    \begin{subfigure}[t]{0.32\textwidth}\includegraphics[width=\textwidth]{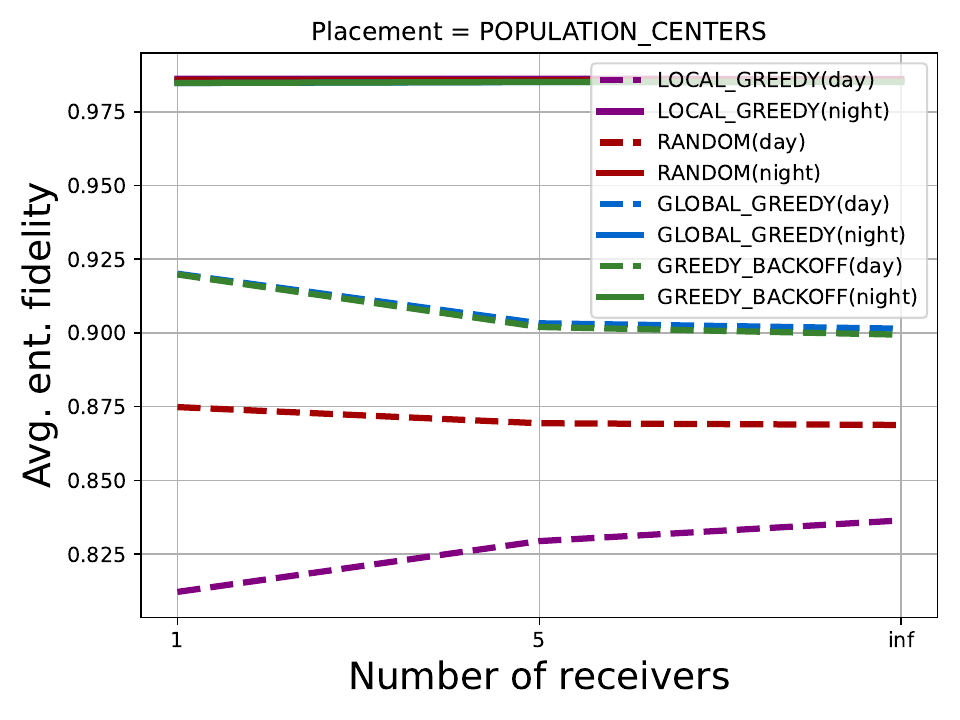}
    \caption{}\end{subfigure}
    \begin{subfigure}[t]{0.32\textwidth}\includegraphics[width=\textwidth]{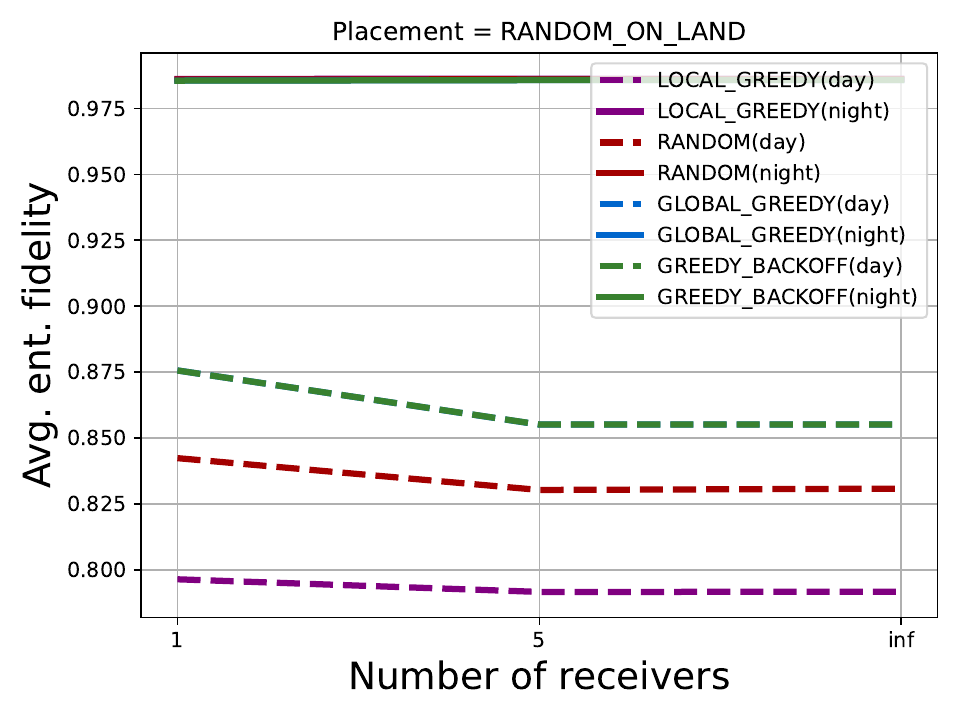}
    \caption{}\end{subfigure}
    \begin{subfigure}[t]{0.32\textwidth}\includegraphics[width=\textwidth]{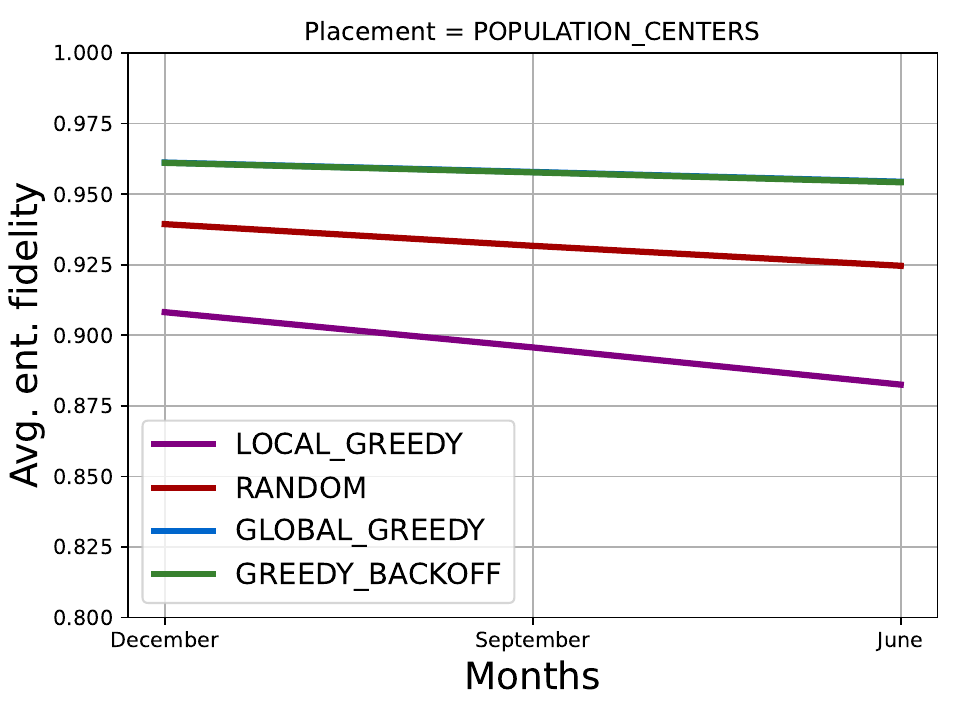}
    \caption{}\end{subfigure}
    % \begin{subfigure}[t]{0.32\textwidth}\includegraphics[width=\textwidth]{new_figs/mean_fidelity_bargraph_algopt_n100_Ti1_Rg1_distpopulation.png}\caption{}\end{subfigure}
    \caption{(a) Satellite altitudes and inclination angles in the Starlink LEO mega-constellation. (b) Maximum number of satellites visible per ground station pair over time. (c) Maximum number of ground station pairs visible per satellite over time. (d) Entanglement distribution rates achieved by different heuristics. (e) Average number of connections per ground station for different heuristics. (f)-(g) Effect of number of receivers on performance. (h)-(i) Longevity of the assigned connections. (j)-(k) Average fidelity vs number of receivers across day time and night time. (l) Effect of different time of the year on entanglement fidelity.}
    \label{fig:all-plots}
\end{figure*} 
% \begin{table}[]
% \begin{tabular}{l|l|l|l|l}
% \hline
% \textbf{Shell}&\textbf{Altitude (Km)} & $\pmb{N_O}$ & $\pmb{N_{SO}}$ & $\pmb{\theta_i}$ (Deg)                                                                \\ \hline
% $1$&$550$&$72$&$22$&52
% \end{tabular}
% \caption{Orbital characteristics of Starlink constellation: $N_O$: \# of orbits, $N_{SO}$: \# of satellites per orbit, $\theta_i:$ Inclination angle of orbit.\textbf{(Parameters to be updated)}}
% \label{table:starlink}
% \end{table}
\subsection{The Starlink constellation}
The Starlink constellation considered in this work consists of $3967$ satellites, distributed across four different orbital inclinations. Figure \ref{fig:all-plots}(a) illustrates the altitudes and orbital inclination angles of these satellites. While the majority orbit at an altitude of $550$ km, there are instances where satellites are positioned at lower altitudes, with a minimum altitude of approximately  $250$ km. The constellation includes four distinct orbital inclination angles: $43, 53, 70,$ and $97$ degrees. At $550$ km altitude, the satellites take around $100$ minutes to complete a full revolution around the earth. To track the positions and trajectories of these satellites, we utilize a two-line element set (TLE) file (available in our code repository). Figure \ref{fig:sat_gs_distributions}(a) displays the spatial distribution of the satellites above Earth's surface. Note that, the density of the satellites is higher around altitudes of approximately $52$ degrees North and South  due to orbital characteristics, whereas density decreases when not between those latitudes.

\subsection{Ground Stations}
\begin{table}[]
\begin{tabular}{l|l|l|l|l|l}
\hline
\textbf{Param}&\textbf{Value} & \textbf{Param}&\textbf{Value} &
\textbf{Param}&\textbf{Value} \\ \hline
$\lambda_s$&$737$ nm&$N_S$&$0.078$&$\tau$&$10^9$\\
$\eta_s$&$0.707$&$\eta_g$&$0.707$&$r_s$&$0.1$ m\\
$r_g$&$1$ m&$t_A$&$5$ km&$\theta_e$&$20$ Deg\\\hline
\end{tabular}
\caption{Simulation Parameters: $\lambda_s, N_S:$ Wavelength and pump power of SPDC source; $\tau:$ Repetition rate of SPDC source; $\eta_s, \eta_g:$ Inefficiencies at satellite transmitters and ground station receivers; $r_s, r_g:$ Radii of circular apertures of the transmitter and receiver telescopes; $t_A:$ Thickness of the atmospheric shell; $\theta_e:$ Elevation angle limit.}
\label{table:params}
\end{table}
We examine two primary placement configurations for $100$ ground stations defined as follows.
\begin{itemize}
    \item \textbf{RANDOM\_ON\_LAND:} In this configuration, ground stations are uniformly placed  at random on land. We first choose a latitude and a longitude uniformly at random. We then resample if the chosen coordinates do not fall on land. Figure \ref{fig:sat_gs_distributions}(b) illustrates the spatial distribution of the ground stations in this configuration.
    \item \textbf{POPULATION\_CENTERS:} Here, we select ground stations corresponding to the $100$ most populous population centers. These centers are chosen from a list maintained at \cite{Youderain_2021}. A portion of the ground station distribution is depicted in Figure \ref{fig:sat_gs_distributions}(c).
\end{itemize}
We assume that for a satellite to be visible by a particular ground station, its elevation angle must be above $\theta_e = 20$ degrees above the horizon. Consequently, each satellite at an altitude of $550$ km has a footprint with an approximate radius of  $1125$ km. This also suggests that ground stations exceeding $2250$ km apart will not be able to establish direct connections via a Starlink satellite positioned at $550$ km above the earth's surface.

\subsection{Simulation Setup}
%\vspace{-0.1cm}
The simulation parameters are outlined in Table \ref{table:params}. To calculate the weights, $w_{ij}$, we use the quadratic free space and exponential atmospheric loss model as detailed in Section \ref{preliminaries}. From this, we compute transmissivities, which in turn, are utilized to determine entanglement distribution rates based on the distances between satellite $i$ and the ground stations in pair $j$, along with an atmospheric thickness of $5$ km above ground level.  We consider three different $24$ hour periods: the two solstices (June 21 and December 22) and one equinox (September 23). We divide each $24$ hour interval into $1$ minute intervals. We divide each day into a night period and a day period. We set the values of detector dark click probability ($P_d$) to $3\times10^{-3}$ and $3\times10^{-7}$ during day and night time respectively \cite{Harney2022}. At the beginning of each time interval, we use the four heuristic algorithms to solve QSSP.  Our simulation is coded in Python 3 \cite{python}, and leverages  third party packages such as the PyEphem \cite{Rhodes_2021}, Matplotlib \cite{Hunter:2007}, Networkx \cite{hagberg2008exploring}, NumPy \cite{harris2020array}, and global-land-mask \cite{todd_karin_2020_4066722}.
%\dt{I would introduce the three day scenarios first.  Then state that each day is divided into one minute intervals over which a scheduling decision is made.}
%\vspace{-1.42cm}
%\subsection{Implementation details}
\subsection{Satellite and ground station pair visibility}
We investigate the visibility of satellites and ground station pairs in Figures \ref{fig:all-plots}(b) and (c). The temporal progression of the peak number of satellites visible to a pair of ground stations is presented in Figure \ref{fig:all-plots}(b). It is interesting to observe that when ground stations are randomly distributed on land, ground station pairs observe fewer satellites compared to when ground stations are placed in highly populous centers, due to the latter resulting in more geographic clusters of ground stations. We plot the maximum number of ground station pairs visible to a satellite in Figure \ref{fig:all-plots}(c). The results demonstrate that the Starlink constellation configuration suits more to population centers even in the quantum case with dual ground station pair visibility requirement. We observe that up to $50$ satellites are visible to a single ground station pair during some of the time slots, while there are times that $40$ ground station pairs are visible to a satellite. This motivates the importance of scheduling, given the large number of potential combinations of satellite to ground station pair assignments.

\begin{figure}
    %\centering
    \includegraphics[width = 3.2in, height = 3in]{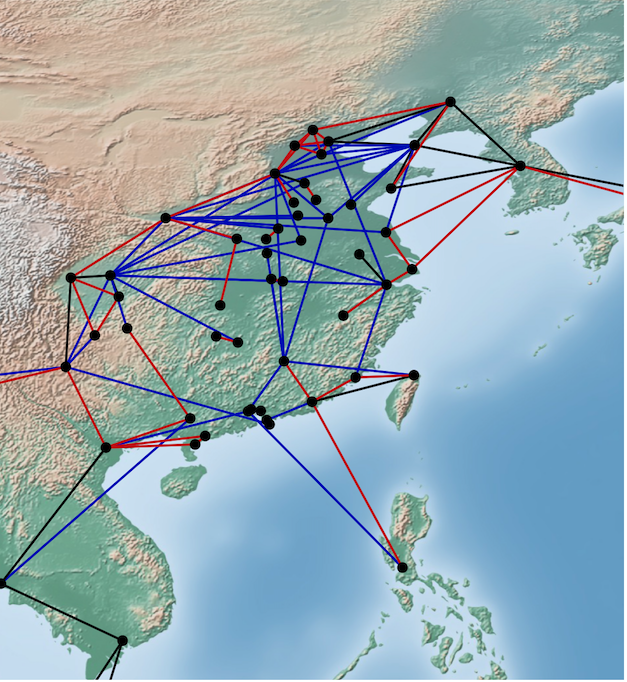}
    \caption{Ground stations in Asia connected with LOCAL\_GREEDY, and GREEDY\_BACKOFF. Red links are the connections assigned only by GREEDY\_BACKOFF, blue links are assigned only by LOCAL\_GREEDY, and black links are assigned by both.}
    \label{fig:asia-map}
\end{figure}

\subsection{Performance of different heuristics}
In Figure \ref{fig:all-plots}(d), we plot the average entanglement distribution rates achieved across different heuristics and ground station placements. We find that rates are lower when ground stations are randomly positioned compared to placements based on population across all heuristic policies. The reason is evident from the visibility plots (Figure \ref{fig:all-plots} (b) and (c)). There are more satellites visible per pair and more ground station pairs visible per satellite in the population center case. So it easier to resolve conflicts, particularly when $R_g =1.$ Additionally, we observe that GLOBAL\_GREEDY, and GREEDY\_BACKOFF perform the best, achieving  nearly identical entanglement distribution rates regardless of ground station placement.

We observe that LOCAL\_GREEDY performs the worst compared to all other algorithms, even falling below RANDOM. Upon further investigation, we observe that LOCAL\_GREEDY tends to prioritize either nearby connections or distant connections, unlike other algorithms (e.g. GREEDY\_BACKOFF) which tend to choose connections of relatively medium distance. This is evident from Figure \ref{fig:asia-map} where we show connections established by LOCAL\_GREEDY and GREEDY\_BACKOFF for a single time interval between ground stations in Asia, when positioned at population centers. LOCAL\_GREEDY tends to choose the best and the worst links, while GREEDY\_BACKOFF opts for average distance connections. This relatively egalitarian assignment in GREEDY\_BACKOFF leads to substantial improvements in entanglement distribution rate over LOCAL\_GREEDY, as entanglement distribution rate is a decreasing convex function of distance.

%RANDOM, (ii) LOCAL\_GREEDY, and (iii) GLOBAL\_GREEDY GREEDY\_BACKOFF
%We find that 

\subsection{Effect of number of receivers}
We study the sensitivity of the performance of different heuristics algorithms to the number of receivers for population center based placement in Figure \ref{fig:all-plots}(f). We observe up to $70\%$ improvement in performance when transitioning from $R_g =1$ to $R_g = 5.$ When $R_g = 5,$ the performance is very similar to that of a scenario with an infinite number of receivers ($R_g = \infty$). This is evident from the number of connections per ground station plot with $R_g = \infty$ shown in Figure \ref{fig:all-plots}(e). When ground stations are placed at population centers, they are form clusters, such as in East Asia and Europe. With more ground station pairs competing for the same limited number of satellites within range of that cluster, most ground stations do not need more than five receivers. Note that, in Figure \ref{fig:all-plots}(e), a  good portion of the ground stations in LOCAL\_GREEDY are assigned to more than five receivers. This explains the performance gap for LOCAL\_GREEDY between $R_g =5$ and $R_g = \infty$ in Figure \ref{fig:all-plots}(f). 

We also explore the impact of the number of receivers on  performance of RANDOM\_ON\_LAND placement, as shown in Figure \ref{fig:all-plots}(g). We observe that the performance gap between $R_g =1$ and  $R_g = 5$ is further increased. Moreover, the performance gap between $R_g = 5$ and $R_g=\infty$ is also increased. %This suggests that there is minimal loss in performance when the  number of receivers is reduced, particularly when ground stations are randomly placed.

% \subsection{Connection ordinality}
% In Figure \ref{fig:all-plots}(h), we plot the distribution of connections allocated to each ground station by each algorithm, assuming no limit on the number of connections a single ground station can establish ($R_g = \infty$). This serves as a rough measure of fairness among ground stations. The width of a violin in the plot shows the mean frequency of connection choices to each ground station assigned by the given algorithm. 
% For instance, consider a ground station located in Boston. Assume that GLOBAL\_GREEDY assigns to Boston five connections which have the first, third, and ninth highest entanglement distribution rates among all possible combinations Boston can form a link with. Then Boston's contribution to the width of the violin plot for that timeslot  would occur at $y$ values $1, 3,$ and $9$.

% Here, we observe that LOCAL\_GREEDY tends to choose some connections that may be of lower priority for the involved ground stations. LOCAL\_GREEDY randomly orders the ground stations and then chooses the best available connection for each ground station in the order. The remaining satellites for ground stations near the end of the random order are unlikely to be provided with the connections with the highest entanglement distribution rates.

\subsection{Longevity of connections}
We track the duration of connections between given ground station pair and a given satellite for each heuristic algorithm in the $R_g=1$ setting, illustrated in Figures \ref{fig:all-plots} (h) and (i). We note that the best-performing heuristic algorithm, GREEDY\_BACKOFF, maintains more connections for multiple timeslots than other algorithms do, with the worst-performing algorithm LOCAL\_GREEDY, maintaining fewer connections for multiple timeslots than any other algorithm. This suggests that GREEDY\_BACKOFF's performance may improve further when accounting for any potential overhead incurred when re-aligning satellite transmitters and receivers to form new connections.

\subsection{Effect of time of day and year}
We plot the average fidelity of the distributed entanglements as a function of number of receivers, during day and night times as shown in Figures \ref{fig:all-plots}(j) and (k). We observe that increasing the number of receivers results in a slight decrease in average fidelity across most heuristics. LOCAL\_GREEDY has the lowest fidelity, while GREEDY\_BACKOFF and GLOBAL\_GREEDY perform the best. The day time fidelities are considerably lower compared to nighttime fidelities across all heuristics and ground station placements. Infidelities caused during the daytime are primarily due to the presence of background photons emitted from sun, in addition to infidelities caused by two-photon pairs emitted at the source during both day and night. 

We also observe that daytime fidelities are lower when ground stations are randomly placed on land versus when they are placed in population centers. This is due to the shorter average distances between ground station pairs in the latter case. We plot the achieved entanglement fidelity across different times of the year for different heuristics in Figure \ref{fig:all-plots}(l). When ground stations are placed in populous centers, the majority of them are in the northern hemisphere. So the average fidelity is smaller for the June solstice due to the extended daylight hours. Maximum fidelity is observed during the December solstice, which has the longest night in northern hemisphere.

% We plot the fidelities for different algorithms in Figure \ref{} ().
%\subsection{Effect of weather}

% \section{Related Literature}
% \input{rel-lit}

\section{Conclusion and Future directions}
In this work, we analyzed QSSP for a LEO  satellite mega-constellation with a large set of ground stations. Via NP-hardness reduction, we proved that the general version of QSSP is intractable. We ran extensive simulations to explore the effectiveness of four heuristic  algorithms to find a schedule for the Starlink mega-constellation. Going forward, we would like to simulate more diverse hardware, such as satellites with more transmitters, satellites capable of establishing inter-satellite links, or even satellites with onboard quantum memories. We would also like to investigate the constellation design problem for quantum satellites which would involve determining the optimal arrangement and configuration of satellites in space for entanglement distribution.

%Going forward, we would like to develop a scheduling algorithm that would prioritize some measure of fairness. This would help discourage malicious ground stations and diversify the chosen assignments, leading to a more connected quantum internet. 

%We also discovered that algorithms that favor consistent medium-distance connections over short and long distance ones perform substantially better.

% \section{ACTION ITEMS}
% \input{action_items}

\bibliographystyle{IEEEtran}
\bibliography{bib}

\begin{thebibliography}{10}
\providecommand{\url}[1]{#1}
\csname url@samestyle\endcsname
\providecommand{\newblock}{\relax}
\providecommand{\bibinfo}[2]{#2}
\providecommand{\BIBentrySTDinterwordspacing}{\spaceskip=0pt\relax}
\providecommand{\BIBentryALTinterwordstretchfactor}{4}
\providecommand{\BIBentryALTinterwordspacing}{\spaceskip=\fontdimen2\font plus
\BIBentryALTinterwordstretchfactor\fontdimen3\font minus \fontdimen4\font\relax}
\providecommand{\BIBforeignlanguage}[2]{{%
\expandafter\ifx\csname l@#1\endcsname\relax
\typeout{** WARNING: IEEEtran.bst: No hyphenation pattern has been}%
\typeout{** loaded for the language `#1'. Using the pattern for}%
\typeout{** the default language instead.}%
\else
\language=\csname l@#1\endcsname
\fi
#2}}
\providecommand{\BIBdecl}{\relax}
\BIBdecl

\bibitem{PhysRevLett.67.661}
\BIBentryALTinterwordspacing
A.~K. Ekert, ``Quantum cryptography based on bell's theorem,'' \emph{Phys. Rev. Lett.}, vol.~67, pp. 661--663, Aug 1991. [Online]. Available: \url{https://link.aps.org/doi/10.1103/PhysRevLett.67.661}
\BIBentrySTDinterwordspacing

\bibitem{Degen_2017}
\BIBentryALTinterwordspacing
C.~Degen, F.~Reinhard, and P.~Cappellaro, ``Quantum sensing,'' \emph{Reviews of Modern Physics}, vol.~89, no.~3, jul 2017. [Online]. Available: \url{https://doi.org/10.1103\%2Frevmodphys.89.035002}
\BIBentrySTDinterwordspacing

\bibitem{Ma_2012}
\BIBentryALTinterwordspacing
X.-S. Ma, T.~Herbst, T.~Scheidl, D.~Wang, S.~Kropatschek, W.~Naylor, B.~Wittmann, A.~Mech, J.~Kofler, E.~Anisimova, V.~Makarov, T.~Jennewein, R.~Ursin, and A.~Zeilinger, ``Quantum teleportation over 143 kilometres using active feed-forward,'' \emph{Nature}, vol. 489, no. 7415, pp. 269--273, sep 2012. [Online]. Available: \url{https://doi.org/10.1038\%2Fnature11472}
\BIBentrySTDinterwordspacing

\bibitem{Pirandola_2017}
\BIBentryALTinterwordspacing
S.~Pirandola, R.~Laurenza, C.~Ottaviani, and L.~Banchi, ``Fundamental limits of repeaterless quantum communications,'' \emph{Nature Communications}, vol.~8, no.~1, apr 2017. [Online]. Available: \url{https://doi.org/10.1038\%2Fncomms15043}
\BIBentrySTDinterwordspacing

\bibitem{D_r_1999}
\BIBentryALTinterwordspacing
W.~Dür, H.-J. Briegel, J.~I. Cirac, and P.~Zoller, ``Quantum repeaters based on entanglement purification,'' \emph{Physical Review A}, vol.~59, no.~1, pp. 169--181, jan 1999. [Online]. Available: \url{https://doi.org/10.1103\%2Fphysreva.59.169}
\BIBentrySTDinterwordspacing

\bibitem{panigrahy2022optimal}
N.~K. Panigrahy, P.~Dhara, D.~Towsley, S.~Guha, and L.~Tassiulas, ``Optimal entanglement distribution using satellite based quantum networks,'' 2022.

\bibitem{Lu2022}
C.~Y. Lu, Y.~Cao, C.~Z. Peng, and J.~W. Pan, ``{Micius quantum experiments in space},'' \emph{Reviews of Modern Physics}, vol.~94, no.~3, 2022.

\bibitem{Gundogan2021}
M.~G{\"{u}}ndoğan, J.~S. Sidhu, V.~Henderson, L.~Mazzarella, J.~Wolters, D.~K. Oi, and M.~Krutzik, ``{Proposal for space-borne quantum memories for global quantum networking},'' \emph{npj Quantum Information}, vol.~7, no.~1, pp. 1--11, 2021.

\bibitem{DeForgesdeParny2023}
L.~{de Forges de Parny}, O.~Alibart, J.~Debaud, S.~Gressani, A.~Lagarrigue, A.~Martin, A.~Metrat, M.~Schiavon, T.~Troisi, E.~Diamanti, P.~G{\'{e}}lard, E.~Kerstel, S.~Tanzilli, and M.~{Van Den Bossche}, ``{Satellite-based quantum information networks: use cases, architecture, and roadmap},'' \emph{Communications Physics}, vol.~6, no.~1, pp. 1--17, 2023.

\bibitem{Khatri21:Spooky}
S.~Khatri, A.~J. Brady, R.~A. Desporte, M.~P. Bart, and J.~P. Dowling, ``Spooky action at a global distance: analysis of space-based entanglement distribution for the quantum internet,'' \emph{npj Quantum Inf}, vol.~7, no.~4, 2021.

\bibitem{Spacex}
N.~K. Panigrahy, P.~Dhara, D.~Towsley, S.~Guha, and L.~Tassiulas, ``Spacex starlink,'' 2017.

\bibitem{Telesat}
A.~F. M. O. M.~A. AUTHORIZATION, ``Spacex starlink,'' 2018.

\bibitem{Amazon}
A.~Boyle, ``Amazon to offer broadband access from orbit with 3,236 satellite ‘project kuiper’ constellation,'' 2019.

\bibitem{dhara2022heralded}
P.~Dhara, S.~J. Johnson, C.~N. Gagatsos, P.~G. Kwiat, and S.~Guha, ``Heralded multiplexed high-efficiency cascaded source of dual-rail entangled photon pairs using spontaneous parametric down-conversion,'' \emph{Physical Review Applied}, vol.~17, no.~3, p. 034071, 2022.

\bibitem{Chen23}
\BIBentryALTinterwordspacing
K.~C. Chen, P.~Dhara, M.~Heuck, Y.~Lee, W.~Dai, S.~Guha, and D.~Englund, ``Zero-added-loss entangled-photon multiplexing for ground- and space-based quantum networks,'' \emph{Phys. Rev. Appl.}, vol.~19, p. 054029, May 2023. [Online]. Available: \url{https://link.aps.org/doi/10.1103/PhysRevApplied.19.054029}
\BIBentrySTDinterwordspacing

\bibitem{Karp1972}
\BIBentryALTinterwordspacing
R.~M. Karp, \emph{Reducibility among Combinatorial Problems}.\hskip 1em plus 0.5em minus 0.4em\relax Boston, MA: Springer US, 1972, pp. 85--103. [Online]. Available: \url{https://doi.org/10.1007/978-1-4684-2001-2\_9}
\BIBentrySTDinterwordspacing

\bibitem{Letchford04}
A.~N. Letchford, G.~Reinelt, and D.~O. Theis, ``A faster exact separation algorithm for blossom inequalities,'' \emph{Proceedings of IPCO, LNCS 3064}, pp. 196--205, 2004.

\bibitem{Anstee87}
R.~P. Anstee, ``A polynomial algorithm for b-matchings: An alternative approach,'' \emph{InformationProcessing Letters}, vol.~24, no.~3, pp. 153--157, 1987.

\bibitem{Youderain_2021}
\BIBentryALTinterwordspacing
C.~Youderain, ``World cities database,'' Jun 2021. [Online]. Available: \url{https://simplemaps.com/data/world-cities}
\BIBentrySTDinterwordspacing

\bibitem{Harney2022}
\BIBentryALTinterwordspacing
C.~Harney, A.~I. Fletcher, and S.~Pirandola, ``{End-To-End Capacities of Hybrid Quantum Networks},'' \emph{Physical Review Applied}, vol.~18, no.~1, p.~1, 2022. [Online]. Available: \url{https://doi.org/10.1103/PhysRevApplied.18.014012}
\BIBentrySTDinterwordspacing

\bibitem{python}
\BIBentryALTinterwordspacing
{Python Core Team}, \emph{{Python: A dynamic, open source programming language}}, {Python Software Foundation}, 2019. [Online]. Available: \url{https://www.python.org/}
\BIBentrySTDinterwordspacing

\bibitem{Rhodes_2021}
\BIBentryALTinterwordspacing
B.~Rhodes, Jun 2021. [Online]. Available: \url{https://rhodesmill.org/pyephem/index.html}
\BIBentrySTDinterwordspacing

\bibitem{Hunter:2007}
J.~D. Hunter, ``Matplotlib: A 2d graphics environment,'' \emph{Computing in Science \& Engineering}, vol.~9, no.~3, pp. 90--95, 2007.

\bibitem{hagberg2008exploring}
A.~Hagberg, P.~Swart, and D.~S~Chult, ``Exploring network structure, dynamics, and function using networkx,'' Los Alamos National Lab.(LANL), Los Alamos, NM (United States), Tech. Rep., 2008.

\bibitem{harris2020array}
\BIBentryALTinterwordspacing
C.~R. Harris, K.~J. Millman, S.~J. van~der Walt, R.~Gommers, P.~Virtanen, D.~Cournapeau, E.~Wieser, J.~Taylor, S.~Berg, N.~J. Smith, R.~Kern, M.~Picus, S.~Hoyer, M.~H. van Kerkwijk, M.~Brett, A.~Haldane, J.~F. del R{\'{i}}o, M.~Wiebe, P.~Peterson, P.~G{\'{e}}rard-Marchant, K.~Sheppard, T.~Reddy, W.~Weckesser, H.~Abbasi, C.~Gohlke, and T.~E. Oliphant, ``Array programming with {NumPy},'' \emph{Nature}, vol. 585, no. 7825, pp. 357--362, Sep. 2020. [Online]. Available: \url{https://doi.org/10.1038/s41586-020-2649-2}
\BIBentrySTDinterwordspacing

\bibitem{todd_karin_2020_4066722}
T.~Karin, ``{toddkarin/global-land-mask: Release of version 1.0.0},'' Oct 2020.

\end{thebibliography}

\end{document}